\documentclass[twocolumn,showpacs,preprintnumbers,amsmath,amssymb]{revtex4}
\usepackage{graphicx}% Include figure files
\usepackage{dcolumn}% Align table columns on decimal point
\usepackage{bm}% bold math

\begin{document}

\preprint{APS/123-QED}

\title{Experimental realization of Lieb-Mattis plateau in a quantum spin chain}

\author{H. Yamaguchi$^{1}$, T. Okita$^{1}$, Y. Iwasaki$^{1}$, Y. Kono$^{1}$, N. Uemoto$^{1}$, \\Y. Hosokoshi$^{1}$, T. Kida$^{2}$, T. Kawakami$^{3}$, A. Matsuo$^{4}$, and M. Hagiwara$^{2}$}
%\author{H. Yamaguchi$^{1}$, M. Okada$^1$, Y. Kono$^2$, S. Kittaka$^2$, T. Sakakibara$^2$, T. Okabe$^1$, Y. Iwasaki$^1$, and Y. Hosokoshi$^1$}
%\author{Second Author}
%\email{yamaguchi@p.s.osakafu-u.ac.jp}
\affiliation{
$^1$Department of Physical Science, Osaka Prefecture University, Osaka 599-8531, Japan\\ 
$^2$Center for Advanced High Magnetic Field Science (AHMF), Graduate School of Science, Osaka University, Osaka 560-0043, Japan\\
$^3$Department of Chemistry, Osaka University, Toyonaka, Osaka 560-0043, Japan\\
$^4$Institute for Solid State Physics, the University of Tokyo, Chiba 277-8581, Japan\\
}

%\affiliation{
Second institution and/or address\\
This line break forced% with \\

\date{\today}% It is always \today, today,

\begin{abstract}
We present a mixed spin-(1/2, 5/2) chain composed of a charge-transfer salt (4-Br-$o$-MePy-V)FeCl$_4$. 
We observe the entire magnetization curve up to saturation, which exhibits a clear Lieb-Mattis magnetization plateau and subsequent quantum phase transition towards the gapless Luttinger-liquid phase.
The observed magnetic behavior is quantitatively explained by a mixed spin-(1/2, 5/2) chain model. 
The present results demonstrate a quantum many-body effect based on quantum topology and provide a new stage in the search for topological properties in condensed matter physics.
\end{abstract}

\maketitle
The one-dimensional (1D) spin chain is the simplest quantum spin model in which spins are linearly arranged. 
Variations of spin size and coupling in 1D spin chains cause a variety of quantum many-body phenomena through strong quantum fluctuations. Haldane's conjecture in 1983 came as a great surprise to the condensed matter physics community. 
He predicted that 1D spin chains would have fundamentally different properties depending on the value of the spins~\cite{haldane}. 
Subsequent experimental and theoretical research has demonstrated that the Heisenberg antiferromagnetic (AF) chain with integer spins has an energy gap between the nonmagnetic ground state and the first excited state, whereas that with half-integer spins has no energy gap. 
The topological ground state of the Haldane chain can be described by a valence-bond picture~\cite{AKLT}, where each integer spin is considered as a collection of $S$ = 1/2 and forms a singlet state between $S$ = 1/2 spins on the different sites.

Mixed spin chains with antiferromagnetically coupled alternating $S$ = 1/2 and $S$ ${\textgreater}$ 1/2 have attracted much attention as a research field for examining topological properties in quantum spin systems.
The Leib-Mattis (LM) theorem explains that the ground state of the mixed spin-(1/2,$S$) Heisenberg chain is a ferrimagnet with a value of $S$-1/2\cite{LM}. 
Furthermore, the topological argument by the Oshikawa-Yamanaka-Affleck criterion~\cite{oshikawa} predicts the appearance of a quantized magnetization plateau at ($S$-1/2)/($S$+1/2). 
Therefore, the mixed spin-(1/2,5/2) chain, which we present in this work, is expected to exhibit a 2/3 magnetization plateau~\cite{cal9,cal2}. 
A particular feature of the LM ferrimagnet is the presence of gapless ferromagnetic and gapped AF spin-wave excitations~\cite{cal2,cal1,cal3,cal4,cal6}. 
The AF gap disappears when a magnetic field applied, and a quantum phase transition to the Luttinger-liquid (LL) phase is expected to appear corresponding to the end of the magnetization plateau and accompanied by a singular square root behavior of the magnetization~\cite{cal9,cal2,cal5}.

From an experimental perspective, several candidates for mixed spin-(1/2,$S$) Heisenberg chains have been reported more than 20 years ago. 
Bimetallic coordination compounds with 3$d$ transition metals are typical examples~\cite{exp1,exp4, exp5,exp6,exp8, exp9}. Further, several molecular materials composed of organic radical and Mn$^{2+}$ ion are expected to form mixed spin-(1/2, 5/2) chains~\cite{exp2,exp3}. 
Although several compounds forming the mixed spin chain show a tendency towards a magnetization plateau, those behaviors are not distinguishable from the thermal paramagnetic state and do not provide conclusive evidence for the LM plateau associated with an energy gap~\cite{exp1, exp5, exp9}. 
Considering the strong AF couplings in the candidate compounds, it is difficult to observe the entire predicted LM plateau using actual high-field measurements.

%PLR用に新たに付け加えた要旨
In this Letter, we report the first observation of a LM plateau in a mixed spin-(1/2, 5/2) chain.
We successfully synthesized single crystals of a verdazyl-based salt (4-Br-$o$-MePy-V)FeCl$_4$ [4-Br-$o$-MePy-V = 3-(5-bromo-1-methylpyridinium-2-yl)-1,5-diphenylverdazyl] and our $ab$ $initio$ molecular orbital (MO) calculations indicated the formation of a mixed spin-(1/2, 5/2) with slightly alternating interactions. 
The magnetization curve exhibits a clear LM magnetization plateau and subsequent quantum phase transition towards the gapless LL phase. 
We quantitatively explain the observed magnetic behavior in terms of the mixed spin-(1/2, 5/2) chain.

%実験方法
We synthesized (4-Br-$o$-MePy-V)FeCl$_4$ using a conventional procedure~\cite{procedure} and prepared an iodide salt of the radical cation (4-Br-$o$-MePy-V)I using a reported procedure for salts with similar chemical structures~\cite{mukai}. 
1-butyl-3-methylimidazolium tetrachloroferrate (239 mg, 0.71 mmol) was slowly added to a solution of (4-Br-$o$-MePy-V)I (149 mg, 0.28 mmol) in 10 ml of ethanol and stirred for 40 min. 
The dark-green solid (4-Br-$o$-MePy-V)FeCl$_4$ was then separated by filtration. 
The dark-green residue was recrystallized using mixed solvent of acetonitrile and methanol.
The crystal structure was determined on the basis of intensity data collected using a Rigaku AFC-8R Mercury CCD RA-Micro7 diffractometer. 
High-field magnetization measurement in pulsed magnetic fields of up to approximately 53 T was conducted using a nondestructive pulse magnet.
The magnetic susceptibility was measured using a commercial SQUID magnetometer (MPMS-XL, Quantum Design).
The experimental result was corrected for the diamagnetic contributions calculated by Pascal's method.
The specific heat was measured using a commercial calorimeter (PPMS, Quantum Design) by using a thermal relaxation method.
Considering the isotropic nature of organic radical systems, all experiments were performed using small randomly oriented single crystals.
$Ab$ $initio$ MO calculations were performed using the UB3LYP method with the basis set 6-31$G$($d$,$p$) in the Gaussian 09 program package. 
For the estimation of intermolecular magnetic interaction, we applied our evaluation scheme that have been studied previously~\cite{MOcal}.
The QMC code is based on the directed loop algorithm in the stochastic series expansion representation~\cite{QMC33}. 
The calculations for the mixed spin-(1/2,5/2) Heisenberg chain were performed for $N$ = 128 under the periodic boundary condition, where $N$ denotes the system size. 
It was confirmed that there is no significant size-dependent effect. 
All calculations were carried out using the ALPS application~\cite{QMC34,QMC35}.

%結晶構造
The molecular structure of  (4-Br-$o$-MePy-V)FeCl$_4$ is shown in Fig.1(a). 
The verdazyl radical 4-Br-$o$-MePy-V and the FeCl$_4$ anion have spin-1/2 and 5/2, respectively. 
The crystallographic parameters are as follows: monoclinic, space group $P$2$_1$/$c$, $a$ = 7.676(7) $\rm{\AA}$, $b$ = 19.425(15) $\rm{\AA}$, $c$ = 16.028(13) $\rm{\AA}$, $\beta$ = 94.962(16)$^{\circ}$, V = 2381(3) $\rm{\AA}^3$, $Z$ = 4, $R$ = 0.0451, and $R_{\rm{w}}$ = 0.0994. 
The 4-Br-$o$-MePy-V and FeCl$_4$ form a 1D structure along the $c$ axis, as shown in Fig. 1(b). 
There are two types of short contacts between the N atom in the central verdazyl ring and the Cl atom, which are 4.04 $\rm{\AA}$ and 3.80 $\rm{\AA}$, respectively. 
$Ab$ $initio$ MO calculations were performed in order to evaluate the corresponding exchange interactions between the spins on 4-Br-$o$-MePy-V and FeCl$_4$ molecules. 
They are evaluated as $J/k_{\rm{B}}$ = 3.9 K and $J'/k_{\rm{B}}$ = 3.3 K, which are defined in the Heisenberg spin Hamiltonian given by $\mathcal {H} = J{\sum^{}_{i}}\textbf{{\textit s}}_{2i-1}{\cdot}\textbf{{\textit S}}_{2i}+J'{\sum^{}_{i}}\textbf{{\textit S}}_{2i}{\cdot}\textbf{{\textit s}}_{2i+1},$, where $\textbf{{\textit s}}$ and $\textbf{{\textit S}}$ are the spin-1/2 and 5/2 operators, respectively. 
The $J$ and $J'$ have very close values and from a mixed spin-(1/2, 5/2) chain with a slight alternation ($J'/J$ = 0.85), as shown in Fig. 1(c). 
We note that intermolecular interactions between organic radical and 3$d$ transition metal anions tend to be underestimated. 
From our previous study~\cite{FeCl4}, the actual values in verdazyl-based salts are expected to be approximately three times larger than those of the MO evaluation, i.e.,  $J/k_{\rm{B}}$ $\simeq$ 12 K and $J'/k_{\rm{B}}$ $\simeq$ 10 K. 
The MO calculations also indicated two types of interchain interactions. 
However, the effective interchain couplings are sufficiently small to be ignored in the current discussion of the mixed spin chain, as described in the following analysis of magnetization.

%磁化
Figure 2(a) shows the magnetization curve at 1.5 K. 
We observe a clear 2/3 magnetization plateau between approximately 2 and 30 T and subsequent gapless behavior. 
The magnetization curve reaches its saturation at approximately 40 T. 
Considering the isotropic $g$ value of $\sim$2.0 resulting from radical and Fe$^{3+}$ ion in this compound, the saturation value of 5.7${\mu}_{\rm{B}}$/f.u. indicates that the purity of the radicals and anions is approximately 95$\%$. 
We consider this purity in the following analysis.
The field derivative of the magnetization curve ($dM/dB$) exhibits a double peak structure accompanied by a phase transition towards the gapless LL phase. 
The magnetization curve below the plateau region should correspond to the thermal paramagnetic state stabilized at finite temperatures, as expected from the Mermin-Wagner theorem~\cite{mermin}.

%基底状態の議論
Here, we discuss the ground state of the present mixed spin-(1/2,5/2) chain. We consider the uniform ($J=J'$) and alternating ($J{\textgreater}J'$) types. 
The uniform type is a typical model forming the LM ferrimagnet. 
Its ground state has the total spin $S_{\rm{tot}}$ = 2 per unit cell, which are degenerate at zero-field. 
The degeneracy is lifted by the finite magnetic field, and the eigenstate with $S_{\rm{tot}}^z$ = 2 becomes the lowest in the plateau region, as described in Fig. 2(b). 
The LM theorem is also applied to the alternating type, and thus the ground state is expected to become a similar ferrimagnetic state with $S_{\rm{tot}}^z$ = 2. 
We can understand the situation by the valence bond picture in Fig. 2(b), where the two spin-1/2 particles on the different sites form a nonmagnetic singlet dimer through $J$. 
We calculated the magnetization curves of the mixed spin-(1/2,5/2) chain using the quantum Monte Carlo (QMC) method. 
The calculated result for the uniform type is confirmed to be the same as that of the previous works~\cite{cal9,cal2}, and is shown in Fig. 2(a). 
We obtained good agreement between the experiment and calculation by using $J/k_{\rm{B}}$ = 9.0 K. 
The calculated result reproduces not only the 2/3 magnetization plateau but also the field-induced LL phase. 
Further, the calculated $dM/dB$ reproduces the double peak structure qualitatively. 
The slight differences in the low-field thermal paramagnetic and LL phases are considered to originate from the interchain interactions. 
Below the plateau phase, the effective spin-2 state fully polarizes at only approximately 2 T, which indicates the interchain couplings are much weaker than the intrachain ones. 
These weak interchain couplings are understood by the valence bond picture of the alternating type in Fig. 2(b). 
The MO calculations indicate that the dominate interchain interactions are between $s$=1/2 radicals. 
However, the $s$=1/2 spins can be regarded as forming a nonmagnetic singlet dimer. 
The effective interactions between the remaining spins are caused through the triplet excited states of the singlet dimers and become much weaker than the MO evaluations~\cite{second1,second2,second3}.

%α依存性
We examined the $\alpha$ = $J'/J$ dependence of the magnetization curve. 
Figures 3(a) and 3(b) show the magnetization curve and $dM/dB$ in the vicinity of the gapless phase for representative values of $\alpha$, respectively. 
The value of $J$ was determined for each $\alpha$, so as to reproduce the central magnetic field in the gapless phase. Hence, the following values were obtained: $J/k_{\rm{B}}$ = 10.6 K ($\alpha$ = 0.7), $J/k_{\rm{B}}$ = 11.9 K ($\alpha$ = 0.5), and $J/k_{\rm{B}}$ = 13.5 K ($\alpha$ = 0.3). 
The magnetization curve increases relatively rapidly as the $\alpha$ decreases. 
The corresponding differences appear in $dM/dB$, where the shape of the peak dramatically changes with decreasing $\alpha$. The decrease of $\alpha$ corresponds to a reduction of one-dimensionality towards a complete dimer without an intermediate LL phase for $\alpha$ = 0. 
We confirm that the uniform type for $\alpha$ = 1 is closest to the experimental result, and the $\alpha$ dependence becomes almost indistinguishable for 0.7 ${\textless}$ $\alpha$ ${\textless}$ 1. 
Accordingly, the actual value of $\alpha$ in the present model is considered to be between 0.7 and 1, which is consistent with the MO evaluation.

%磁化率
Figure 4(a) shows the temperature dependence of the magnetic susceptibility ($\chi$ = $M/H$) at 0.1 T. 
We observe an anomalous change at 5.0 K, which indicates that a phase transition to the long-range order (LRO) occurs at $T_{\rm{N}}$ = 5.0 K owing to the small but finite interchain interactions. 
In the higher-temperature region, the value of ${\chi}T$ exhibits a rounded minimum at approximately 24 K, as shown in the inset of Fig. 4(a). 
This behavior is a typical characteristic of mixed spin chains~\cite{exp4,exp5,exp6,exp8}. 
We then calculated ${\chi}T$ for the uniform mixed spin-(1/2,5/2) chain with $J/k_{\rm{B}}$ = 9.0 K by using the QMC method, obtaining a good agreement in the temperature region above $T_{\rm{N}}$.

%比熱
The experimental result for the specific heat at zero-field clearly exhibits a $\lambda$-type sharp peak associated with the phase transition to the LRO, which is consistent with the discontinuous change in the magnetic susceptibility, as shown in Fig. 4(b). 
We have not observed the expected double peak and $\sqrt{T}$ behavior in the low-temperature region~\cite{cal2} due to the phase transition behavior. 
Because the experimental behavior below $T_{\rm{N}}$ exhibits almost $T$-linear dependence, the gapless ferromagnetic dispersion is considered to change to an AF linear dispersion owing to the effect of the AF interchain interactions. 
In magnetic fields, the phase transition temperature decreases with increasing fields and almost disappears above 2 T, which is consistent with the formation of the plateau phase with a spin gap.

%まとめ
In summary, we have succeeded in synthesizing single crystals of the verdazyl-based salt (4-Br-$o$-MePy-V)FeCl$_4$
Our $ab$ $initio$ MO calculations indicate the formation of a mixed spin-(1/2, 5/2) chain. 
We observe the entire magnetization curve up to saturation, which exhibits a clear Lieb-Mattis magnetization plateau and subsequent quantum phase transition towards the gapless LL phase.
The magnetic susceptibility and specific heat also indicate behavior consistent with the formation of the Lieb-Mattis plateau.
The present results demonstrate a quantum many-body effect based on the Lieb-Mattis theorem and will stimulate studies on topological properties in mixed spin chains.

\begin{figure*}[t]
\begin{center}
\includegraphics[width=36pc]{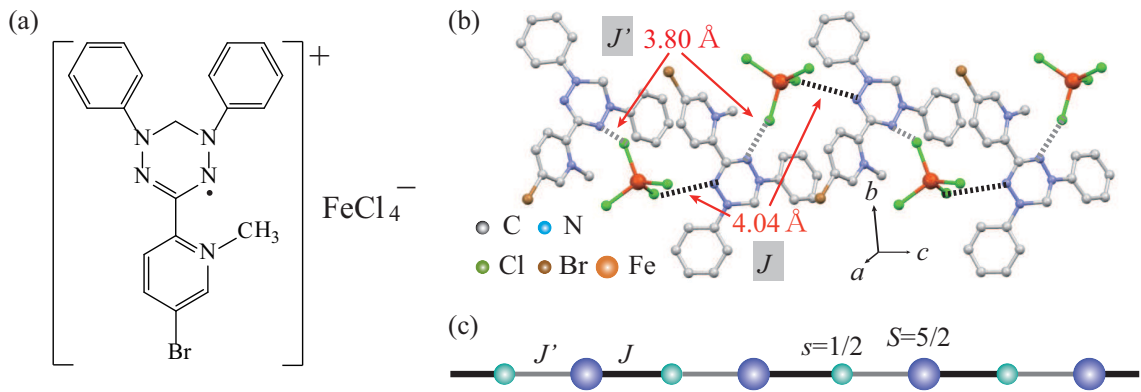}
\caption{(color online) (a) Molecular structures of (4-Br-$o$-MePy-V)FeCl$_4$. (b) Crystal structure forming a 1D chain along the $c$ axis. Hydrogen atoms are omitted for clarity. The broken lines indicate two types of N-Cl short contact associated with $J$ and $J'$. (c) Mixed spin-(1/2,5/2) chain composed of $s$ = 1/2 on the radical and $S$ =5/2 on the FeCl$_4$ anion through $J$ and $J'$.}\label{f1}
\end{center}
\end{figure*}

\begin{figure*}[t]
\begin{center}
\includegraphics[width=36pc]{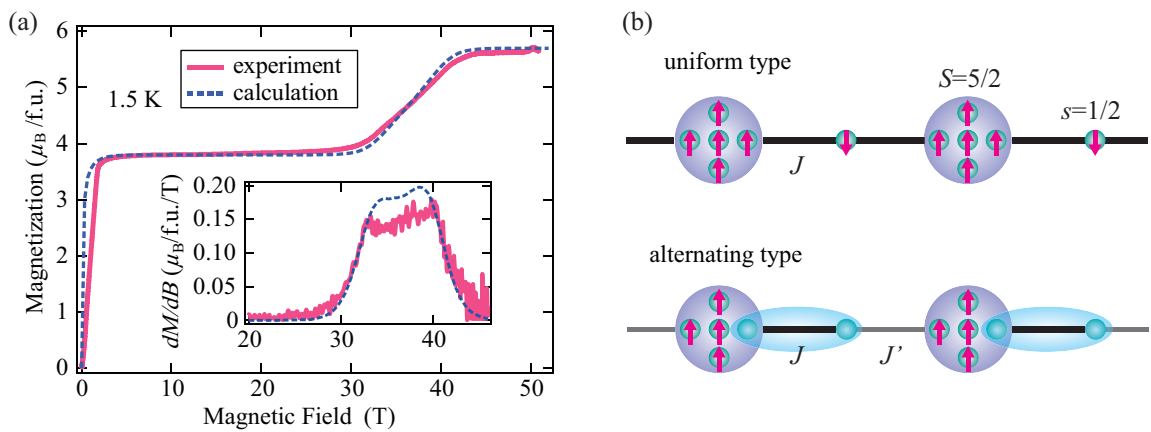}
\caption{(color online) (a) Magnetization curve of (4-Br-$o$-MePy-V)FeCl$_4$ at 1.5 K. The inset shows the field derivative of the magnetization curve ($dM/dB$). The broken lines represent the calculated results for the mixed spin-(1/2,5/2) chain ($J=J'$, $\alpha$ = 1) using the QMC method. (b) Valence bond pictures for the LM plateau on the two types of mixed spin-(1/2,5/2) chain, i.e., uniform ($J=J'$) and alternating ($J{\textgreater}J'$) types. $S$ = 5/2 spins are described by five $s$ = 1/2 spins. The ovals represent valence bond singlet pairs of two $s$ = 1/2 spins, $\frac{1}{\sqrt{2}}({\uparrow}{\downarrow}-{\downarrow}{\uparrow})$.}\label{f2}
\end{center}
\end{figure*}

\begin{figure}[t]
\begin{center}
\includegraphics[width=18pc]{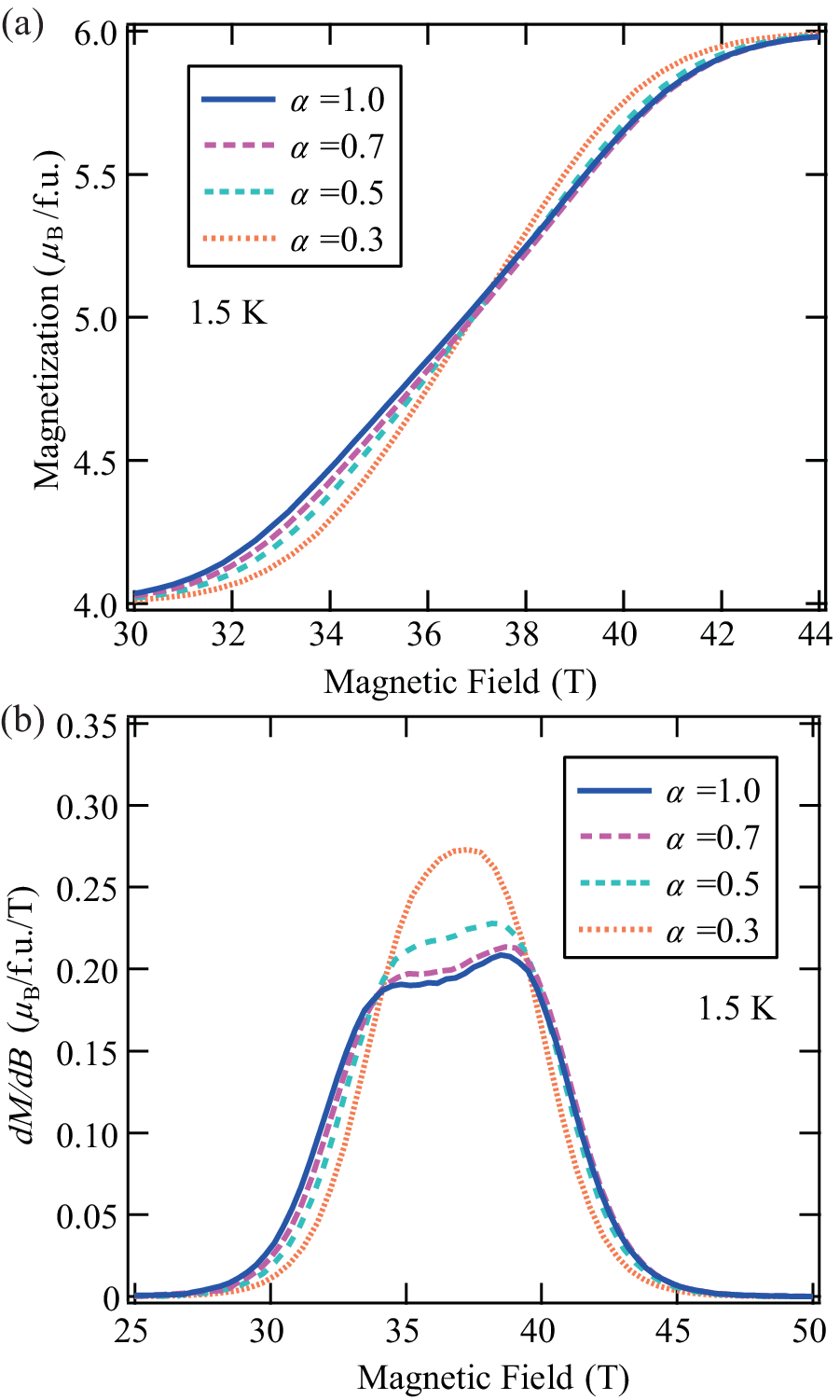}
\caption{(color online) (a) Calculated magnetization curves and (b) their field derivative at 1.5 K for the mixed spin-(1/2,5/2) chain with various values of $\alpha$ by using QMC method. The magnetic field range extends the gapless region where the $\alpha$ dependence is relatively remarkable. }\label{f3}
\end{center}
\end{figure}

\begin{figure}[t]
\begin{center}
\includegraphics[width=18pc]{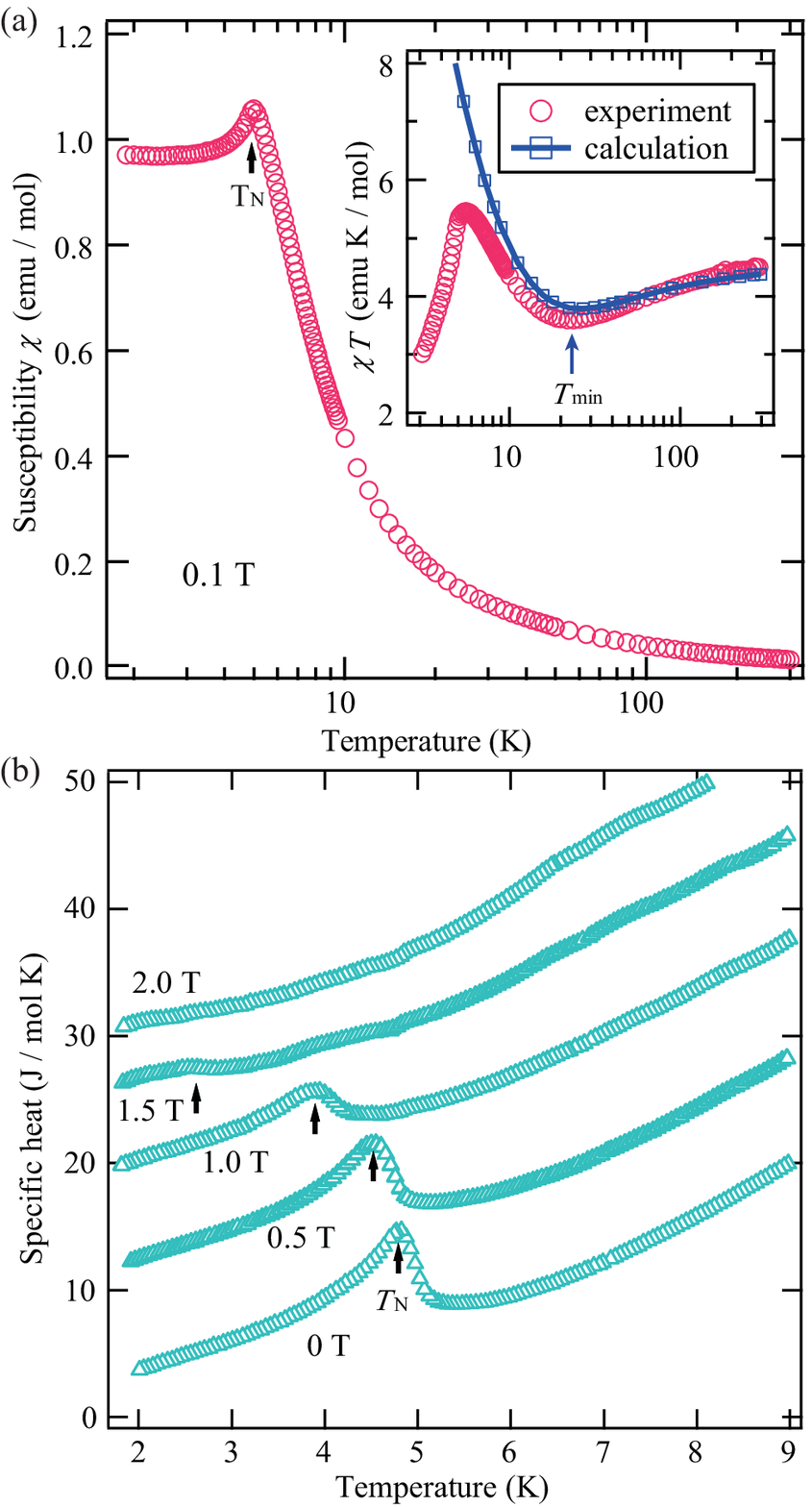}
\caption{(color online) (a) Temperature dependence of magnetic susceptibility ($\chi$ = $M/H$) at 0.1 T. The arrow indicates the phase transition temperature $T_{\rm{N}}$. The inset shows the temperature dependence of $\chi T$, where the position of the rounded minimum is marked with Tmin. The solid line with open squares represents calculated result for the mixed spin-(1/2,5/2) chain ($J=J'$, $\alpha$ = 1) by using QMC method. (b) Temperature dependence of the specific heat at various magnetic fields. The arrows indicate the phase transition temperatures $T_{\rm{N}}$. For clarity, the values for 0.5, 1.0, 1.5, and 2.0 T have been shifted up by 8.6, 16, 22, and 27 J mol$^{-1}$ K, respectively.}\label{f4}
\end{center}
\end{figure}

We thank T. Tonegawa, K. Okamoto, and S. C. Furuya for valuable discussions. This research was partly supported by KAKENHI (No. 17H04850). A part of this work was carried out at the Center for Advanced High Magnetic Field Science in Osaka University under the Visiting Researcher's Program of the Institute for Solid State Physics, the University of Tokyo, and the Institute for Molecular Science.

%%%%%%%%%%%%%%%%%%%%%%%%%%%%%%%%%%%%%%%%%%%%%%%%%%%%%%%%%%%%%%
%%%%%%%%%%%%

\end{document}